\documentclass[eqsecnum,prd,showpacs,amsmath,aps,nofootinbib]
{revtex4}
\usepackage{amsmath}
\usepackage{amsfonts}
\usepackage{amsmath}
\usepackage{amssymb}

\begin{document}
\title{Non-linear constitutive equations for gravitoelectromagnetism}

\author{Steven Duplij}
\email[E-mail: ]{sduplij@gmail.com}
\affiliation{Theory Group, Nuclear Physics Laboratory, V.N. Karazin
Kharkov National University, Svoboda Sq. 4, Kharkov 61077, Ukraine}

\author{Elisabetta Di Grezia}
\email[E-mail: ]{digrezia@na.infn.it}
\affiliation{Istituto Nazionale di Fisica Nucleare, Sezione di
Napoli, Complesso Universitario di Monte S. Angelo, 
Via Cintia Edificio 6, 80126 Napoli, Italy}

\author{Giampiero Esposito}
\email[E-mail: ]{gesposit@na.infn.it}
\affiliation{Istituto Nazionale di Fisica Nucleare, Sezione di
Napoli, Complesso Universitario di Monte S. Angelo, 
Via Cintia Edificio 6, 80126 Napoli, Italy}

\author{Albert Kotvytskiy} 
\email[E-mail: ]{kotvytskiy@gmail.com}
\affiliation{Department of Physics, V.N. Karazin Kharkov National
University, Svoboda Sq. 4, Kharkov 61077, Ukraine}

\begin{abstract}
This paper studies non-linear constitutive equations for
gravitoelectromagnetism. Eventually, the problem is solved of
finding, for a given particular solution of the gravity-Maxwell
equations, the exact form of the corresponding non-linear 
constitutive equations. 
\end{abstract}


\maketitle

\section{Introduction}

Over the past decade, a description of non-linear classical
electrodynamics and Yang-Mills theory has been considered in the
literature \cite{gol/sht1, gol/sht2, dup/gol/sht}, with the hope of
being able to extend it to a broader framework, including gauge
theories of gravity \cite{Neeman1998} and quantum gravity 
\cite{Espo2011}.

However, no explicit calculation had been performed,
and the formulation remained too general for the physics community 
to be able to appreciate its potentialities. For this purpose, as a 
first step, we here consider the gravito-electromagnetism 
in the weak-field approximation
(following, e.g., \cite{clark}). Recall the standard Maxwell equations in SI
units \cite{jackson}
\begin{align}
\operatorname{curl}\mathbf{E}  &  =-\dfrac{\partial\mathbf{B}}{\partial
t},\ \ \ \ \ \ \operatorname{div}\mathbf{B}=0,\nonumber\\
\operatorname{curl}\mathbf{H}  &  =\dfrac{\partial\mathbf{D}}{\partial
t}+\mathbf{j},\ \ \ \operatorname{div}\mathbf{D}=\rho, \label{max3a}
\end{align}
where $\mathbf{E}$ is the electric field, $\mathbf{B}$ is the magnetic field,
$\rho$ is charge density, $\mathbf{j}$ is electric current density. In the
linear case
\begin{equation}
\mathbf{B}=\mu_{0}\mathbf{H},\ \ \ \mathbf{D}=\varepsilon_{0}\mathbf{E,}
\label{be}
\end{equation}

In the non-linear case these equations can be presented in the form
\cite{fus/sht/ser}
\begin{align}
\mathbf{D}  &  =M\left(  I_{1},I_{2}\right)  \mathbf{B}
+\dfrac{1}{c^{2}}N\left(  I_{1},I_{2}\right)  \mathbf{E},\nonumber\\
\mathbf{H}  &  =N\left(  I_{1},I_{2}\right)  \mathbf{B}-M\left(  I_{1}
,I_{2}\right)  \mathbf{E}, \label{con1}
\end{align}
where the invariants are ($F_{\mu \nu}$ being the electromagnetic
field tensor, with Hodge dual ${ }^{*}F^{\mu \nu}$)
\begin{equation}
I_{1}={1\over 2}F_{\mu \nu}F^{\mu \nu}
=\mathbf{B}^{2}-\dfrac{1}{c^{2}}\mathbf{E}^{2},
\ I_{2}=-{c\over 4}F_{\mu \nu}{ }^{*}F^{\mu \nu}
=\mathbf{B} \cdot\mathbf{E.\ }
\end{equation}
Their gravitational analogues in SI are
\begin{align}
\operatorname{curl}\mathbf{E}_{g}  &  
=-\dfrac{\partial\mathbf{B}_{g}}{\partial t},
\ \ \ \ \ \ \operatorname{div}\mathbf{B}_{g}=0,
\label{maxg0}\\
\operatorname{curl}\mathbf{B}_{g}  &  =\dfrac{1}{c^{2}}\dfrac{\partial
\mathbf{E}_{g}}{\partial t}+\dfrac{1}{\varepsilon_{g}c^{2}}\mathbf{j}
_{g},\ \ \ \operatorname{div}\mathbf{E}_{g}=\dfrac{1}{\varepsilon_{g}}\rho
_{g}, \label{maxg}
\end{align}
where $\mathbf{E}_{g}$ is the static gravitational field (conventional
gravity, also called gravitoelectric for the sake of analogy), $\mathbf{B}
_{g}$ is the gravitomagnetic field, $\rho_{g}$ is mass density, $\mathbf{j}
_{g}$ is mass current density, $G$ is the gravitational constant,
$\varepsilon_{g}$ is the gravity permittivity (analog of $\varepsilon_{0})$.
Here
\begin{equation}
\varepsilon_{g}=-\dfrac{1}{4\pi G},\ \ \ \ \ \mu_{g}=-\dfrac{4\pi G}{c^{2}},
\label{em}
\end{equation}
are the gravitational permittivity and permeability, respectively.

The main idea is to introduce analogues of $\mathbf{H}$ and $\mathbf{D}$ to
write (\ref{maxg0}) and (\ref{maxg}) 
in the Maxwell form for $4$ fields in SI as
\begin{align}
\operatorname{curl}\mathbf{E}_{g}  &  =-\dfrac{\partial\mathbf{B}_{g}%
}{\partial t},\ \ \ \ \ \ \operatorname{div}\mathbf{B}_{g}=0,\label{maxgg0}\\
\operatorname{curl}\mathbf{H}_{g}  &  =\dfrac{\partial\mathbf{D}_{g}}{\partial
t}+\mathbf{j}_{g},\ \ \ \operatorname{div}\mathbf{D}_{g}=\rho_{g}.
\label{maxgg}
\end{align}
In the linear-gravity case
\begin{align}
\mathbf{D}_{g}  &  =\varepsilon_{g}\mathbf{E}_{g},\label{de}\\
\mathbf{B}_{g}  &  =\mu_{g}\mathbf{H}_{g},\label{bh}\\
\varepsilon_{g}\mu_{g}  &  =\dfrac{1}{c^{2}}. \label{mc}
\end{align}
Note now that the linear-gravity case 
(\ref{de})--(\ref{mc}) corresponds to weak approximation 
and some special case of
gravitational field configuration. We generalize it 
to non-linear case which can describe
other configurations and non-weak fields, as in (\ref{con1}), by
\begin{align}
\mathbf{D}_{g}  &  =M_{g}\left(  I_{g1},I_{g2}\right)  \mathbf{B}_{g}
+\dfrac{1}{c^{2}}N_{g}\left(  I_{g1},I_{g2}\right)  \mathbf{E}_{g}
,\label{dg}\\
\mathbf{H}_{g}  &  =N_{g}\left(  I_{g1},I_{g2}\right)  \mathbf{B}_{g}
-M_{g}\left(  I_{g1},I_{g2}\right)  \mathbf{E}_{g}, \label{hg}
\end{align}
where the invariants are
\begin{equation}
I_{g1}=\mathbf{B}_{g}^{2}-\dfrac{1}{c^{2}}\mathbf{E}_{g}^{2},\ I_{g2}
=\mathbf{B}_{g}\cdot\mathbf{E}_{g}\mathbf{.\ } \label{ig}
\end{equation}

The gravity-Maxwell equations (\ref{maxgg0})--(\ref{maxgg}) together with the
non-linear gravity-constitutive equations (\ref{dg})-(\ref{hg}) can give
a non-linear electrodynamics formulation of gravity 
(or at least some particular instances of this construction).

\section{Linear gravito-electromagnetic waves}

The gravity-Maxwell equations for gravito-electromagnetic waves (far from
sources) are
\begin{align}
\operatorname{curl}\mathbf{E}_{g} &  =-\dfrac{\partial\mathbf{B}_{g}}{\partial
t},\operatorname{div}\mathbf{B}_{g}=0,\\
\operatorname{curl}\mathbf{H}_{g} &  =\dfrac{\partial\mathbf{D}_{g}}{\partial
t},\operatorname{div}\mathbf{D}_{g}=0
\end{align}
with generic values of permittivity and permeability (\ref{em}). Then
\begin{align}
\operatorname{curl}\mathbf{E}_{g} &  =-\mu_{g}\dfrac{\partial\mathbf{H}_{g}
}{\partial t},\operatorname{div}\mathbf{H}_{g}=0,\\
\operatorname{curl}\mathbf{H}_{g} &  =\varepsilon_{g}\dfrac{\partial
\mathbf{E}_{g}}{\partial t},\operatorname{div}\mathbf{E}_{g}=0.
\end{align}
We differentiate the first equation with respect to time: $\operatorname{curl}
\dfrac{\partial}{\partial t}\mathbf{E}_{g}=-\mu_{g}\dfrac{\partial
^{2}\mathbf{H}_{g}}{\partial t^{2}}$ $\Rightarrow\dfrac{1}{\varepsilon_{g}
}\operatorname{curl}\left(  \operatorname{curl}\mathbf{H}_{g}\right)
=-\mu_{g}\dfrac{\partial^{2}\mathbf{H}_{g}}{\partial t^{2}}.$ Since
$\operatorname{curl}\left(  \operatorname{curl}\mathbf{H}_{g}\right)
=\operatorname{grad}(\operatorname{div}\mathbf{H}_{g})-\Delta\mathbf{H}
_{g}=-\Delta\mathbf{H}_{g}$, then
\begin{equation}
\Delta\mathbf{H}_{g}=\varepsilon_{g}\mu_{g}\dfrac{\partial^{2}\mathbf{H}_{g}
}{\partial t^{2}}.
\end{equation}

By analogy, from the second equation $\operatorname{curl}\dfrac{\partial
}{\partial t}\mathbf{H}_{g}=\varepsilon_{g}\dfrac{\partial^{2}\mathbf{E}_{g}
}{\partial t^{2}}$ $\Rightarrow$ $\dfrac{-1}{\mu_{g}}\operatorname{curl}
\left(  \operatorname{curl}\mathbf{E}_{g}\right)  =\varepsilon_{g}
\dfrac{\partial^{2}\mathbf{E}_{g}}{\partial t^{2}}$. Hence we get the wave
equation for $\mathbf{E}_{g}$,
\begin{equation}
\Delta\mathbf{E}_{g}=\varepsilon_{g}\mu_{g}\dfrac{\partial^{2}\mathbf{E}_{g}
}{\partial t^{2}}.
\end{equation}

\section{Nonlinear gravito-electromagnetic waves}

The differences begin with the constitutive equations (\ref{dg})--(\ref{hg}).
For simplicity put first $M_{g}=0$. Then
\begin{align}
\mathbf{D}_{g} &  =\dfrac{N}{c^{2}}\mathbf{E}_{g},\\
\mathbf{B}_{g} &  =\dfrac{1}{N}\mathbf{H}_{g}
\end{align}
where $N\equiv N_{g}\left(  I_{g1},I_{g2}\right)  $. The Maxwell equations
become (hereafter the dots denote time derivatives)
\begin{align}
\operatorname{curl}\mathbf{E}_{g} &  =-\left(  \dfrac{1}{N}\right)  ^{\bullet
}\mathbf{H}_{g}-\dfrac{1}{N}\dfrac{\partial\mathbf{H}_{g}}{\partial
t},\label{1}\\
\operatorname{div}\left(  \dfrac{1}{N}\mathbf{H}_{g}\right)   &
=\mathbf{H}_{g}\operatorname{grad}\left(  \dfrac{1}{N}\right)  +\dfrac{1}
{N}\operatorname{div}\left(  \mathbf{H}_{g}\right)  =0,\label{2}\\
\operatorname{curl}\mathbf{H}_{g} &  =\dfrac{\dot{N}}{c^{2}}\mathbf{E}
_{g}+\dfrac{N}{c^{2}}\dfrac{\partial\mathbf{E}_{g}}{\partial t},\label{3}\\
\operatorname{div}\left(  \dfrac{N}{c^{2}}\mathbf{E}_{g}\right)   &
=\mathbf{E}_{g}\operatorname{grad}\left(  \dfrac{N}{c^{2}}\right)  +\dfrac
{N}{c^{2}}\operatorname{div}\left(  \mathbf{E}_{g}\right)  =0.\label{4}
\end{align}

Take derivative of (\ref{1}) with respect to time and get
\begin{equation}
\operatorname{curl}\dfrac{\partial}{\partial t}\mathbf{E}_{g}=-\left(
\dfrac{1}{N}\right)  ^{\bullet\bullet}\mathbf{H}_{g}-2\left(  \dfrac{1}
{N}\right)  ^{\bullet}\dfrac{\partial\mathbf{H}_{g}}{\partial t}-\dfrac{1}
{N}\dfrac{\partial^{2}\mathbf{H}_{g}}{\partial t^{2}}.
\end{equation}
From (\ref{3}) it follows $\dfrac{\partial\mathbf{E}_{g}}{\partial t}
=\dfrac{c^{2}}{N}\operatorname{curl}\mathbf{H}_{g}-\dfrac{\dot{N}}
{N}\mathbf{E}_{g}$. Then we get
\begin{equation}
\operatorname{curl}\left(  \dfrac{c^{2}}{N}\operatorname{curl}\mathbf{H}
_{g}-\dfrac{\dot{N}}{N}\mathbf{E}_{g}\right)  =-\left(  \dfrac{1}{N}\right)
^{\bullet\bullet}\mathbf{H}_{g}-2\left(  \dfrac{1}{N}\right)  ^{\bullet}
\dfrac{\partial\mathbf{H}_{g}}{\partial t}-\dfrac{1}{N}\dfrac{\partial
^{2}\mathbf{H}_{g}}{\partial t^{2}}.
\end{equation}
The left-hand side here is
\begin{align*}
& \operatorname{curl}\left(  \dfrac{c^{2}}{N}\operatorname{curl}\mathbf{H}
_{g}-\dfrac{\dot{N}}{N}\mathbf{E}_{g}\right)  \\
& =\operatorname{grad}\dfrac{c^{2}}{N}\times\operatorname{curl}\mathbf{H}
_{g}+\dfrac{c^{2}}{N}\operatorname{grad}\operatorname{div}\mathbf{H}
_{g}-\dfrac{c^{2}}{N}\Delta\mathbf{H}_{g}-\dfrac{\dot{N}}{N}
\operatorname{curl}\mathbf{E}_{g}-\operatorname{grad}\dfrac{\dot{N}}{N}
\times\mathbf{E}_{g}.
\end{align*}
From (\ref{2}) we get $\operatorname{div}\left(  \mathbf{H}_{g}\right)
=-N\mathbf{H}_{g}\operatorname{grad}\left(  \dfrac{1}{N}\right)  \neq0$. Thus,
the non-linear analogue of the wave equation is
\begin{align}
&  \operatorname{grad}\dfrac{c^{2}}{N}\times\operatorname{curl}\mathbf{H}
_{g}+\dfrac{c^{2}}{N}\operatorname{grad}\operatorname{div}\mathbf{H}
_{g}-\dfrac{c^{2}}{N}\Delta\mathbf{H}_{g}-\dfrac{\dot{N}}{N}
\operatorname{curl}\mathbf{E}_{g}-\operatorname{grad}\dfrac{\dot{N}}{N}
\times\mathbf{E}_{g}\\
&  =-\left(  \dfrac{1}{N}\right)  ^{\bullet\bullet}\mathbf{H}_{g}-2\left(
\dfrac{1}{N}\right)  ^{\bullet}\dfrac{\partial\mathbf{H}_{g}}{\partial
t}-\dfrac{1}{N}\dfrac{\partial^{2}\mathbf{H}_{g}}{\partial t^{2}}.
\end{align}
Note that if $N={\rm const}$, then we obtain the usual wave equation
\begin{equation}
\Delta\mathbf{H}_{g}=\dfrac{1}{c^{2}}\dfrac{\partial^{2}\mathbf{H}_{g}
}{\partial t^{2}}.
\end{equation}

Take now the constitutive equations in the form
\begin{align}
\mathbf{D}_{g} &  =M\mathbf{B}_{g}+\dfrac{N}{c^{2}}\mathbf{E}_{g},\\
\mathbf{H}_{g} &  =N\mathbf{B}_{g}-M\mathbf{E}_{g},
\end{align}
where $N,M$ are constants. In absence of sources, the Maxwell equations
become
\begin{align}
\operatorname{curl}\mathbf{E}_{g} &  =-\dfrac{\partial\mathbf{B}_{g}}{\partial
t},\ \ \ \operatorname{div}\mathbf{B}_{g}=0,\\
\operatorname{curl}\mathbf{H}_{g} &  =\dfrac{\partial\mathbf{D}_{g}}{\partial
t},\ \ \ \operatorname{div}\mathbf{D}_{g}=0.
\end{align}
If we express the Maxwell equations through $\mathbf{E}_{g}$ and
$\mathbf{B}_{g}$, the second pair of equations become
\begin{equation}
\operatorname{curl}\mathbf{H}_{g}   
=\dfrac{\partial\mathbf{D}_{g}}{\partial t} \Longrightarrow
N\operatorname{curl}\mathbf{B}_{g}-M\operatorname{curl}\mathbf{E}_{g}  
=M\dfrac{\partial\mathbf{B}_{g}}{\partial t}+\dfrac{N}{c^{2}}\dfrac
{\partial\mathbf{E}_{g}}{\partial t}.
\end{equation}

Since $\operatorname{curl}\mathbf{E}_{g}=-\dfrac{\partial\mathbf{B}_{g}
}{\partial t}$, we get
\begin{equation}
\operatorname{curl}\mathbf{B}_{g}=\dfrac{1}{c^{2}}\dfrac{\partial
\mathbf{E}_{g}}{\partial t}.
\end{equation}
The second equation, $\operatorname{div}\mathbf{D}_{g}=0$, reduces to
$M\operatorname{div}\mathbf{B}_{g}+\dfrac{N}{c^{2}}\operatorname{div}
\mathbf{E}_{g}=0$. Since $\operatorname{div}\mathbf{B}_{g}=0,$ we get
\begin{equation}
\operatorname{div}\mathbf{E}_{g}=0.
\end{equation}
Thus, using constitutive equations with constant $M$ and $N$ we have Maxwell
equations in terms of $\mathbf{B}_{g}$ and $\mathbf{E}_{g}$, i.e.
\begin{align}
\operatorname{curl}\mathbf{E}_{g} &  =-\dfrac{\partial\mathbf{B}_{g}}{\partial
t},\ \ \ \operatorname{div}\mathbf{B}_{g}=0,\\
\operatorname{curl}\mathbf{B}_{g} &  =\dfrac{1}{c^{2}}\dfrac{\partial
\mathbf{E}_{g}}{\partial t},\ \ \ \operatorname{div}\mathbf{E}_{g}=0.
\end{align}
At this stage, we get the wave equations in the standard way. The time
derivative of the first equation yields $\operatorname{curl}\dfrac{\partial
}{\partial t}\mathbf{E}_{g}=-\dfrac{\partial^{2}\mathbf{B}_{g}}{\partial
t^{2}}$ $\Rightarrow c^{2}\operatorname{curl}\left(  \operatorname{curl}
\mathbf{B}_{g}\right)  =-\dfrac{\partial^{2}\mathbf{B}_{g}}{\partial t^{2}}$.
Since $\operatorname{curl}\left(  \operatorname{curl}\mathbf{B}_{g}\right)
=\operatorname{grad}(\operatorname{div}\mathbf{B}_{g})-\Delta\mathbf{B}
_{g}=-\Delta\mathbf{B}_{g}$, then
\begin{equation}
\Delta\mathbf{B}_{g}=\dfrac{1}{c^{2}}\dfrac{\partial^{2}\mathbf{B}_{g}
}{\partial t^{2}}.
\end{equation}
By analogy $\operatorname{curl}\dfrac{\partial}{\partial t}\mathbf{B}
_{g}=\dfrac{1}{c^{2}}\dfrac{\partial^{2}\mathbf{E}_{g}}{\partial t^{2}}$
$\Rightarrow$ $-\operatorname{curl}\left(  \operatorname{curl}\mathbf{E}
_{g}\right)  =\dfrac{1}{c^{2}}\dfrac{\partial^{2}\mathbf{E}_{g}}{\partial
t^{2}}$, and we get the wave equation for $\mathbf{E}_{g}$,
\begin{equation}
\Delta\mathbf{E}_{g}=\dfrac{1}{c^{2}}\dfrac{\partial^{2}\mathbf{E}_{g}
}{\partial t^{2}}.
\end{equation}
Thus, the gravi-electromagnetic waves $\mathbf{E}_{g}$ and $\mathbf{B}_{g}$
have speed $c$ and do not depend on the constants $M$ and $N$.

\section{Waves and constitutive equations for linear constitutive functions}

Let us consider the constitutive equations (\ref{dg})--(\ref{hg}) as linear
functions of the invariants, i.e.
\begin{align}
M &  =M_{g}\left(  I_{g1},I_{g2}\right)  =a_{m}I_{g1}+b_{m}I_{g2},\\
N &  =N_{g}\left(  I_{g1},I_{g2}\right)  =c^{2}\varepsilon_{g}+a_{n}
I_{g1}+b_{n}I_{g2},
\end{align}
$a_{m},b_{m},a_{n},b_{n}$ being some constants. From all the Maxwell equations
in material media, and in the absence of sources one finds
$\operatorname{curl}\mathbf{H}_{g}=\dfrac{\partial\mathbf{D}_{g}}{\partial t}
$, $\operatorname{curl}\left(  N\mathbf{B}_{g}-M\mathbf{E}_{g}\right)
=\dfrac{\partial}{\partial t}\left(  M\mathbf{B}_{g}+\dfrac{N}{c^{2}
}\mathbf{E}_{g}\right)  $, and $N\operatorname{curl}\mathbf{B}_{g}
-M\operatorname{curl}\mathbf{E}_{g}=M\dfrac{\partial\mathbf{B}_{g}}{\partial
t}+\dfrac{N}{c^{2}}\dfrac{\partial\mathbf{E}_{g}}{\partial t}$. Since
$\operatorname{curl}\mathbf{E}_{g}=-\dfrac{\partial\mathbf{B}_{g}}{\partial
t}$, from the last equation one gets
\begin{equation}
\operatorname{curl}\mathbf{B}_{g}=\dfrac{1}{c^{2}}\dfrac{\partial
\mathbf{E}_{g}}{\partial t}.
\end{equation}
The second equation, $\operatorname{div}\mathbf{D}_{g}=0$, reduces to
$\operatorname{div}\left(  M\mathbf{B}_{g}+\dfrac{N}{c^{2}}\mathbf{E}
_{g}\right)  =0$, or $M\operatorname{div}\mathbf{B}_{g}+\dfrac{N}{c^{2}
}\operatorname{div}\mathbf{E}_{g}=0$. Since $\operatorname{div}\mathbf{B}
_{g}=0$, one gets
\begin{equation}
\operatorname{div}\mathbf{E}_{g}=0.
\end{equation}

\section{Inverse problem of non-linear gravito-electromagnetism}

In electrodynamics the direct solution of the Maxwell equations together with
the non-linear constitutive equations is a 
non-trivial and complicated task even
for simple systems \cite{gol/sht1,gol/sht2}. In previous sections we presented
some very special cases of the non-linear functions $N$ and $M$. Here we
formulate the following inverse problem: if we have some particular solution
of the gravity-Maxwell equations (\ref{maxgg0})--(\ref{maxgg}), can we then
find the exact form of the corresponding non-linear gravity-constitutive
equations (\ref{dg})-(\ref{hg})?

It is natural to consider the case of plane gravitational waves, when the
fields have only one space coordinate. We will show that even in this case one
can have a non-trivial non-linearity. Let us choose $\mathbf{E}_{g}$ and
$\mathbf{B}_{g}$ mutually orthogonal and perpendicular to the direction of
motion
\begin{equation}
\mathbf{E}_{g}=\left(
\begin{array}
[c]{c}
E\\
0\\
0
\end{array}
\right)  ,\ \ \ \mathbf{B}_{g}=\left(
\begin{array}
[c]{c}
0\\
0\\
B
\end{array}
\right)  ,\label{e}
\end{equation}
where $E\equiv E\left(  t,y\right)$, $B\equiv B\left(  t,y\right)  $. 
Now the invariants (\ref{ig}) become
\begin{align}
I_{g1} &  =B^{2}-\dfrac{1}{c^{2}}E^{2}\equiv I,\label{i}\\
I_{g2} &  =0\mathbf{.\ }
\end{align}
The use of the non-linear gravity-constitutive 
equations (\ref{dg})-(\ref{hg}) gives for the other fields
\begin{equation}
\mathbf{D}_{g}=\left(
\begin{array}
[c]{c}
\dfrac{1}{c^{2}}NE\\
0\\
MB
\end{array}
\right)  ,\ \ \ \mathbf{H}_{g}=\left(
\begin{array}
[c]{c}
-ME\\
0\\
NB
\end{array}
\right)  ,
\label{dh}
\end{equation}
where $N\equiv N\left(I\right)$, $M\equiv M\left(I\right)$ are the
sought for gravity-constitutive functions. They depend on $I$ only, because of
Lorentz invariance (see \cite{gol/sht1,gol/sht2}). Inserting the fields
(\ref{e}) and (\ref{dh}) into the gravity-Maxwell equations 
(\ref{maxgg0})--(\ref{maxgg}) without sources gives us $3$ equations
(hereafter, a prime with the corresponding subscript denotes
the first partial derivative with respect to the variable 
in the subscript, while dot denotes time derivative)
\begin{align}
E_{y}'  & =\dot{B},\label{eb}\\
\left(NB\right)_{y}'  & =\dfrac{1}{c^{2}}\left(NE\right)^{\cdot},
\label{nb}\\
\left(ME\right)_{y}'  & =\left(MB\right)  ^{\cdot}.
\label{me}
\end{align} 
Now we take into account 
that the gravity-constitutive functions $N$, $M$
depend only on the invariant $I$ and present (\ref{nb})--(\ref{me}) 
as the differential equations for them
\begin{align}
N_{I}'\left(BI_{y}'-\dfrac{1}{c^{2}}E\dot{I}\right)  +N\left(
B_{y}'-\dfrac{1}{c^{2}}\dot{E}\right) & =0,
\label{ni}\\
M_{I}'\left(EI_{y}'-B\dot{I}\right) & =0,
\label{mi}
\end{align}
where we have exploited the identities
\begin{equation}
N_{y}'=N_{I}' I_{y}', \; M_{y}'=M_{I}' I_{y}',
\end{equation}
\begin{equation}
{\dot N}=N_{I}'{\dot I}, \; {\dot M}=M_{I}'{\dot I}.
\end{equation} 
The second equation (\ref{mi}) can be immediately solved by
\begin{equation}
M\left(  I\right)  =\left\{
\begin{array}
[c]{c}
M_{0}={\rm const},\ \ \ \text{if }EI_{y}'\neq B\dot{I},\\
{\rm arbitrary},\ \ \ \text{if }EI_{y}'=B\dot{I}.
\end{array} \right.  
\label{m1}
\end{equation}
The first equation (\ref{ni}) can be solved if 
\begin{equation}
\lambda \equiv {\left(B_{y}'-{{\dot E}\over c^{2}}\right) \over
\left(BI_{y}'-{E {\dot I}\over c^{2}}\right)}
\end{equation}
depends only on $I$, which is a very special case. 
One then has the differential equation
\begin{equation}
N_{I}'+\lambda\left(I\right)N=0,
\end{equation}
and its solution is
\begin{equation}
N\left(  I\right)  =N_{0}{\rm e}^{-\int\lambda
\left(I\right) {\rm d}I}.
\end{equation}
Otherwise, by using the expressions for $I_{y}'$ and 
$\dot{I}$ from (\ref{i}), i.e.
\begin{equation}
I_{y}'=2BB_{y}'-{2EE_{y}'\over c^{2}}, \;
{\dot I}=2B{\dot B}-{2E{\dot E}\over c^{2}},
\end{equation}
we obtain
\begin{equation}
2N_{I}'\left(B^{2}B_{y}'+\dfrac{1}{c^{4}}E^{2}\dot{E}-\dfrac
{2}{c^{2}}EBE_{y}'\right)  +N\left(B_{y}'-\dfrac{1}{c^{2}}\dot
{E}\right)=0,
\label{2ni}
\end{equation}
where the sum of terms in brackets is not a function of $I$ in general.

Usually, in the wave solutions the dependence of fields on frequency $\omega$
and wave number $k$ is the same, and therefore we can consider the
concrete choice
\begin{equation}
E\left(t,y\right)  =f\left(\varepsilon\omega t+ky\right)\equiv f(X(t,y))  ,
\ \ \ B\left(t,y\right)
=g\left(\varepsilon\omega t+ky\right)\equiv g(X(t,y))  ,
\end{equation}
where $\varepsilon \equiv \pm 1$, with $f$ and $g$ arbitrary smooth
nonvanishing functions. Bearing in mind that 
$$
E_{y}'=f_{X}'X_{y}'=k f_{X}', \; 
B_{y}'=g_{X}'X_{y}'=k g_{X}',
$$
$$
{\dot E}=f_{X}'{\dot X}=\varepsilon \omega f_{X}', \;
{\dot B}=g_{X}'{\dot X}=\varepsilon \omega g_{X}',
$$
our Eq. (\ref{eb}) yields
\begin{equation}
kf_{X}'=\varepsilon \omega g_{X}'.
\end{equation}
Therefore
\begin{equation}
g\left(X\right)  =\dfrac{k}{\varepsilon \omega}f
\left(X\right)  +\alpha,
\end{equation}
where $\alpha$ is a constant, so that both $E$ and $B$ can be
expressed through one function only, i.e. $f$, and the invariant $I$
reads eventually as 
\begin{equation}
I=\dfrac{1}{\omega^{2}}\left(  k^{2}-\dfrac{\omega^{2}}{c^{2}}\right)
f^{2}+2\dfrac{k}{\varepsilon \omega}\alpha f+\alpha^{2}.
\end{equation}
The equations for the gravity-constitutive functions 
take therefore the form
\begin{align}
N_{I}'\left[2I\left(k^{2}-{\omega^{2}\over c^{2}}\right)
+2 {\omega^{2}\over c^{2}} \alpha^{2} \right]
+N \left(k^{2}-{\omega^{2} \over c^{2}}\right)    & =0,
\label{w1}\\
M_{I}'f_{X}'\left[{2f \over \varepsilon \omega}
\left(k^{2}-{\omega^{2}\over c^{2}}\right)+2k \alpha \right]
\alpha & =0,
\label{w2}
\end{align}
having exploited the identities
\begin{equation}
gI_{y}'-{f {\dot I}\over c^{2}}=\left(gf_{y}'
-{f {\dot f}\over c^{2}}\right)
\left[{2f \over \omega^{2}}\left(k^{2}-{\omega^{2}\over c^{2}}\right)
+2{k \over \varepsilon \omega}\alpha \right],
\end{equation}
\begin{equation}
gf_{y}'-{f {\dot f}\over c^{2}}={f_{X}'\over \varepsilon \omega}
\left[f \left(k^{2}-{\omega^{2}\over c^{2}}\right)
+k \varepsilon \omega \alpha \right],
\end{equation}
and, after some cancellations,
\begin{eqnarray}
\; & \; & \left[f \left(k^{2}-{\omega^{2}\over c^{2}}\right)
+k \varepsilon \omega \alpha \right]
\left[{2f \over \omega^{2}}\left(k^{2}-{\omega^{2}\over c^{2}}
\right)+2{k \over \varepsilon \omega}\alpha \right] 
\nonumber \\
&=& 2 \left(k^{2}-{\omega^{2}\over c^{2}}\right)I
+2{\omega^{2}\over c^{2}}\alpha^{2},
\end{eqnarray}
while
\begin{equation}
fI_{y}'-g{\dot I}=(ff_{y}'-g{\dot f})
\left[{2f \over \omega^{2}}
\left(k^{2}-{\omega^{2}\over c^{2}}\right)
+2{k \over \varepsilon \omega}\alpha \right],
\end{equation}
\begin{equation}
ff_{y}'-g{\dot f}=f_{X}'(kf-\varepsilon \omega g)
=-\varepsilon \omega f_{X}' \alpha.
\end{equation}

The results of our analysis now depend on whether or not
$\alpha$ vanishes. Indeed, if $\alpha=0$, $M$ is arbitrary
and hence we obtain the equation
\begin{equation}
\left(k^{2}-{\omega^{2}\over c^{2}}\right)(2IN_{I}'+N)=0,
\end{equation}
which implies that either the dispersion relation
\begin{equation}
k^{2}-{\omega^{2}\over c^{2}}=0
\label{disp}
\end{equation}
holds, with $N$ kept arbitrary, or such a dispersion relation
is not fulfilled, while $N$ is found from the differential equation
\begin{equation}
2IN_{I}'+N=0,
\end{equation}
which is solved by 
\begin{equation}
N(I)={N_{0}\over \sqrt{I}}.
\end{equation}

By contrast, if $\alpha$ does not vanish, $M$ equals a constant 
$M_{0}$, while $N$ solves the more complicated equation
(\ref{w1}). At this stage, to be consistent with the dependence
of $N$ on $I$ only, we have to require again that the dispersion
relation (\ref{disp}) should hold, jointly with $N_{I}'=0$, which
implies the constancy of $N$: $N=N_{0}$. 

\section{Concluding remarks}

We have brought `down to earth' the general program of considering
non-linear constitutive equations for gravitoelectromagnetism, by
solving the problem of finding, for a given solution of the 
gravity-Maxwell equations, the exact form of non-linear constitutive
equations. We look forward to being able to construct other relevant 
examples, as well as being able to re-express our models in the
language of differential forms, which turned out to be very powerful
for general relativity \cite{plebanski,krasnov,capovilla}. 

\acknowledgments

S. Duplij thanks M. Bianchi, J. Gates, G. Goldin, A. Yu. Kirochkin, 
M. Shifman, V. Shtelen, D. Sorokin, A. Schwarz, M. Tonin, A. Vainshtein,  
A. Vilenkin for fruitful discussions. E. Di Grezia and G. Esposito are
grateful to the Dipartimento di Fisica of Federico II University, Naples, for
hospitality and support.


\begin{thebibliography}{}                                                                                               
\bibitem{gol/sht1}
G. A. Goldin and V. M. Shtelen, {\it Phys. Lett. A} {\bf 279} (2001) 321.

\bibitem{gol/sht2}
G. A. Goldin and V. M. Shtelen, {\it J. Phys. A: Math. Gen.} 
{\bf 37} (2004) 10711.

\bibitem{dup/gol/sht}
S. Duplij, G. A. Goldin and V. M. Shtelen,  
{\it J. Phys. A: Math. Gen.} {\bf 41} (2008) 304007.

\bibitem{Neeman1998}
Y. Ne'eman, {\it Acta Phys. Pol.} {\bf 29} (1998) 827.

\bibitem{Espo2011}
G. Esposito, arXiv:1108.3269, in EOLSS Encyclopedia, UNESCO (2011).

\bibitem{clark}
S. J. Clark and R. W. Tucker, {\it Class. Quantum Grav.} {\bf 17}
(2000) 4125.

\bibitem{jackson}
J. D. Jackson, {\it Classical Electrodynamics} (Wiley, New York, 1999).

\bibitem{fus/sht/ser}
W. I. Fushchich, V. M. Shtelen, and N. I. Serov, 
{\it Symmetry Analysis and Exact Solutions of Equations of Non-Linear
Mathematical Physics} (Kluwer, Dordrecht, 1993).

\bibitem{plebanski}
J. F. Plebanski, {\it J. Math. Phys.} {\bf 18} (1977) 2511.

\bibitem{krasnov}
K. Krasnov, {\it Gen. Rel. Grav.} {\bf 43} (2011) 1.

\bibitem{capovilla}
R. Capovilla, J. Dell and T. Jacobson, {\it Class. Quantum Grav.} 
{\bf 8} (1991) 59.

\end{thebibliography}
\end{document}